\documentclass[11pt]{article}
\usepackage[all]{xy}
\usepackage{amssymb,amsmath,
stmaryrd}
\usepackage{diagrams}

\begin{document}
\thispagestyle{plain}
\title{Semantics of Interaction}
\author{Samson Abramsky}

\def\ve{\varepsilon}

\newarrow{Partial}----{harpoon}

\def\proofskip{{\vskip 4pt plus 1pt minus 1pt}}
\def\proofbox{{\hfill\rule{6pt}{6pt}}}
\newenvironment{proof}{{\noindent \sc Proof\/}~}{\proofbox\proofskip}
\newenvironment{openproblem}{{\medskip\noindent \bf Open problem\/}~}{\proofskip}

\newtheorem{DUMREM}{dummy}[section]
\newtheorem{remark}[DUMREM]{{\medskip\noindent \bf Remark\/}~\rm}{\proofskip}
\newtheorem{DUMNOT}{dummy}[section]
\newtheorem{notation}[DUMNOT]{{\medskip\noindent \bf Notation\/}~\rm }{\proofskip}
\newtheorem{DUMEX}{dummy}[section]
\newtheorem{exercise}[DUMEX]{{\medskip\noindent \bf Exercise\/}~\rm}{\proofskip} 
\newtheorem{DUMLEM}{dummy}[section]
\newtheorem{lemma}[DUMLEM]{{\medskip\noindent \bf Lemma\/}~}{\proofskip} 
\newtheorem{DUMPRO}{dummy}[section]
\newtheorem{proposition}[DUMPRO]{{\medskip\noindent \bf Proposition\/}~}{\proofskip} 
\newtheorem{DUMEXAMPLE}{dummy}[section]
\newtheorem{example}[DUMEXAMPLE]{{\medskip\noindent \bf Example\/}~\rm}{\proofskip}

\def\Pref{\mathop{\mathrm{Pref}}\nolimits}
\def\even{\mathop{\mathrm{even}}\nolimits}
\def\odd{\mathop{\mathrm{odd}}\nolimits}
\def\dom{\mathop{\mathrm{dom}}\nolimits}
\def\parity{\mathop{\mathrm{parity}}\nolimits}
\def\state#1{{\ulcorner #1\urcorner}}
\def\draftfigure{\centerline{\large\sf Figure comes here}}

\def\RED{\longrightarrow}
\def\PARA{\: \Vert \:}
\def\DIF{\!\setminus\!}
\def\RES{\restriction}
\def\SIMRED{\stackrel{\sim}{\RED}}  %
\def\ENCan#1{{\langle #1 \rangle}}

\newcommand{\G}{{\cal G}}
\newcommand{\I}{\mbox{I}}
\newcommand{\U}{{\cal U}}
\newcommand{\V}{{\cal V}}

\newcommand{\true}{\mathit{tt}}
\newcommand{\false}{\mathit{ff}}
\newcommand{\eqdef}{\stackrel{\mbox{\tiny def}}{\equiv}}
\newcommand{\Set}{\mbox{\bf Set}}
\newcommand{\one}{\mbox{\bf 1}}

\newcommand{\id}[1]{{\mathtt id}_{#1}}
\newcommand{\Obj}[1]{Obj(#1)}
\newcommand{\lam}[1]{\lambda_{#1}}
\newcommand{\M}[1]{\mbox{M}_{#1}}
\renewcommand{\P}[1]{\mbox{P}_{#1}}
\newcommand{\pop}{\multimap}
\newcommand{\tensor}{\otimes}

\newcommand{\po}{\trianglelefteq}
\newcommand{\veedirect}[2]{{{\bigvee}_{#1}{#2}}}
\newcommand{\cupdirect}[2]{{\bigcup}_{#1}{#2}}
\newcommand{\restrict}{\restriction}

\newcommand{\Ap}[2]{{\mathtt Ap}_{#1,#2}}
\newcommand{\B}{{\Bbb B}}

\newcommand{\K}{{\mathbf K}}
\newcommand{\Gwhf}{{\cal G}^{\mbox{\scriptsize hf}}_{\cal W}}
\newcommand{\GwhfR}{{\cal G}^{\mbox{\scriptsize hf}}_{{\cal W}R}}

\newcommand{\Rel}{\mbox{\em Rel}}
\newcommand{\Span}{\mbox{\em Span}}
\newcommand{\prop}[1]{P#1}
\newcommand{\rel}[1]{\cal{R}_{#1}}
\newcommand{\Cat}{{\cal C}} 



\newcommand{\THEN}{{\,\,\, \; \Rightarrow \; \,\,\,}}
\newcommand{\IFF}{{\,\,\, \; \Leftrightarrow \; \,\,\,}}
\newcommand{\ASET}[1]{\{ #1 \}}
\newcommand{\AND}{\wedge}
\newcommand{\AAND}{\bigwedge}
\newcommand{\OR}{\vee}
\newcommand{\DEFEQ}{\equiv}
\newcommand{\MNTN}[1]{\widehat{{#1}}}
\newcommand{\NUL}{\varepsilon}
\newcommand{\RS}{\upharpoonright}
\newcommand{\ODR}{\trianglelefteq}
\newcommand{\NPRF}{\subseteq^{\mathrm{nepref}}}

\newcommand{\lmerge}{\, \llfloor \,}
\newcommand{\rmerge}{\, \rrfloor \,}
\newcommand{\linimpl}{\multimap}
\newcommand{\proj}{{\upharpoonright}}

\maketitle

\begin{abstract}
The ``classical'' paradigm for denotational semantics models data types
as {\em domains}, {\sl i.e.} structured sets of some kind, and programs as
(suitable) {\em functions} between domains. The semantic universe in which
the denotational modelling is carried out is thus a category with domains
as objects, functions as morphisms, and composition of morphisms given by
function composition.
A sharp distinction is then drawn between denotational and operational
semantics. Denotational semantics is often referred to as ``mathematical
semantics'' because it exhibits a high degree of mathematical structure;
this is in part achieved by the fact that denotational semantics abstracts
away from the dynamics of computation---from time.
By contrast, operational semantics is formulated in terms of the
syntax of the language being modelled; it is highly intensional in
character; and it is capable of expressing the dynamical aspects of
computation.

The classical denotational paradigm has been very successful, but has
some definite limitations. Firstly, fine-structural features of
computation, such as sequentiality, computational complexity, and
optimality of reduction strategies, have either not been captured
at all denotationally, or not in a fully satisfactory fashion.
Moreover, once languages with features beyond the purely
functional are considered, the appropriateness of
modelling programs by functions
is increasingly open to question. Neither concurrency nor
``advanced'' imperative features such as local references have
been captured denotationally in a fully convincing fashion.

This analysis suggests a desideratum of {\em Intensional Semantics},
interpolating between denotational and operational semantics as
traditionally conceived. This should combine the good mathematical
structural properties of denotational semantics with the ability to
capture dynamical aspects and to embody computational intuitions
of operational semantics. Thus we may think of Intensional semantics
as ``Denotational semantics + time (dynamics)'', or as ``Syntax-free
operational semantics''.

A number of recent developments (and, with hindsight, some older ones)
can be seen as contributing to this goal of Intensional Semantics.  We
will focus on the recent work on Game semantics, which has led to some
striking advances in the Full Abstraction problem for PCF and other
programming languages (Abramsky {\em et al.} 1995)
(Abramsky and McCusker 1995) (Hyland and Ong 1995) (McCusker 1996a)
(Ong 1996). 
Our aim is to give a genuinely elementary
first introduction; we therefore present a simplified version of game
semantics, which nonetheless contains most of the essential concepts. The
more complex game semantics in (Abramsky {\em et al.} 1995)
(Hyland and Ong 1995) can be seen as refinements
of what we present.  Some background in category theory, type theory
and linear logic would be helpful in reading these notes; suitable
references are (Crole 1994), (Girard {\em et al.} 1989), (Girard 1995)
(which contain much more than we will actually need).

\end{abstract}


\paragraph{Acknowledgements}
The first version of this note comes from notes
for the course on ``Semantics of Interaction'' given at the CLICS II Summer School,
Cambridge, September 24--29, 1995.
I would like to thank the Edinburgh ``interaction group'' (Kohei Honda,
Paul-Andr\'e Melli\`es, Julo Chroboczek, Jim Laird and Nobuko Yoshida) for
their help in preparing these notes for publication,  Peter Dybjer 
for his comments on a draft version, and Peter Dybjer and Andy Pitts 
for their efforts in organizing the CLiCS summer school and editing 
the present volume.

\tableofcontents

\section*{Notation}

If $X$ is a set, $X^*$ is the set of finite sequences (words, strings)
over $X$.  We use $s$, $t$, $u$, $v$ to denote sequences, and $a$,
$b$, $c$, $d$, $m$, $n$ to denote elements of these sequences.
Concatenation of sequences is indicated by juxtaposition, and we won't
distinguish notationally between an element and the corresponding unit
sequence. Thus $as$ denotes the sequence with first element $a$ and 
tail
$s$.

If $f:X\longrightarrow Y$ then $f^*:X^*\longrightarrow Y^*$ is the 
unique
monoid homomorphism extending $f$.  We write $|s|$ for the length of a
finite sequence, and $s_i$ for the $i$th element of $s$, 
$1\leq i\leq |s|$.

Given a set $S$ of sequences, we write $S^{\mathrm{even}}$,
$S^{\mathrm{odd}}$ for the subsets of even- and odd-length sequences
respectively.

We write $X+Y$ for the disjoint union of sets $X$, $Y$.

If $Y\subseteq X$ and $s\in X^*$, we write $s\restriction Y$ for the
sequence obtained by deleting all elements not in $Y$ from $s$. In 
practice, we use this notation in the context where $X = Y + Z$, and 
by abuse of notation we take $s \restriction Y \in Y^{*}$, {\em i.e.}\ we 
elide the use of injection functions. 

We
write $s\sqsubseteq t$ if $s$ is a prefix of $t$, {\em i.e.}\ $t=su$
for some $u$.

$\Pref(S)$ is the set of prefixes of elements of $S\subseteq X^*$.
$S$ is {\em prefix-closed} if $S=\Pref(S)$.

\section{Game Semantics}

We give a first introduction to game semantics.  We will be concerned
with 2-person games. Why the number 2? The key feature of games,
by comparison with the many extant models of computation (labelled
transition systems, event structures, etc. etc.) is that they provide
an {\em explicit representation of the environment}, and hence model
interaction in an intrinsic fashion. (By contrast, interaction is
modelled in, say, labelled transition systems using some additional
structure, typically a ``synchronization algebra'' on the labels.) 
One-person games would degenerate to transition systems; it seems that
multi-party interaction can be adequately modeled by two-person games,
in much the same way that functions with multiple arguments can be
reduced to one-place functions and tupling.  We will use such games to
model interactions between a System and its Environment.  One of the
players in the game is taken to represent the System, and is referred
to as Player or Proponent; the other represents the Environment and is
referred to as Opponent.  Note that the distinction between System and
Environment and the corresponding designation as Player or Opponent
depend on {\em point of view}:
\begin{quote}
If Tom, Tim and Tony converse in a room, then from Tom's point of
view, he is the System, and Tim and Tony form the Environment; while
from Tim's point of view, he is the System, and Tom and Tony form the
Environment.
\end{quote}
A single `computation' or `run' involving interaction between Player
and Opponent will be represented by a sequence of {\em moves}, made
alternately by Player and Opponent.  We shall adopt the convention
that {\em Opponent always makes the first move}.  This avoids a number
of technical problems which would otherwise arise, but limits what we
can successfully model with games to the {\em negative fragment} of
Intuitionistic Linear Logic. (This is the $\otimes$, $\multimap$, $\&$,
$!$, $\forall$ fragment).

A game specifies the set of possible runs (or `plays').  It can be
thought of as a tree
$$
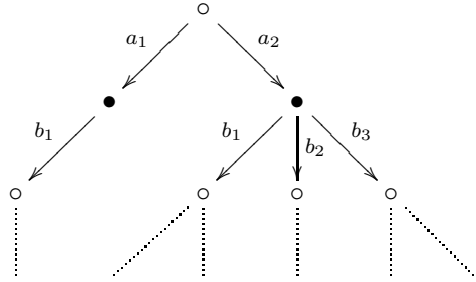
\xymatrix{
  &  &  {\circ}\ar[dl]_{a_1}\ar[dr]^{a_2}\\
  & {\bullet}\ar[dl]_{b_1} & & 
{\bullet}\ar[dl]_{b_1}\ar[d]^{b_2}\ar[dr]^{b_3}\\
  {\circ}\ar@{.}[d] & & 
     {\circ}\ar@{.}[dl]\ar@{.}[d] & {\circ}\ar@{.}[d] & 
       {\circ}\ar@{.}[d]\ar@{.}[dr]\\
  & & & & & & 
}$$
where hollow nodes~$\circ$ represent positions where Opponent is to
move; solid nodes~$\bullet$ positions where Player is to move; and the
arcs issuing from a node are labelled with the moves which can be made
in the position represented by that node.

Formally, we define a game $G$ to be a structure 
  $(M_G,\lambda_G,P_G)$, where
\begin{itemize}
\item $M_G$ is the set of {\em moves} of the game;
\item $\lambda_G : M_G\longrightarrow \{P,O\}$ is a labelling function
  designating each move as by Player or Opponent;
\item $P_G\subseteq^{\mathrm{nepref}}M_G^{\mathrm{alt}}$, {\em i.e.}\
  $P_G$ is a non-empty, prefix-closed subset of $M_G^{\mathrm{alt}}$, 
the
  set of alternating sequences of moves in $M_G$.
\end{itemize}

More formally, $M_G^{\mathrm{alt}}$ is the set of all $s\in M_G^*$ 
such
that
$$\begin{array}{lcl}
\forall i:1\leq i\leq |s|
 &       &\even(i)\Longrightarrow \lambda_G(s_i)=P \\
 &\wedge &\odd(i) \Longrightarrow \lambda_G(s_i)=O
\end{array}$$
{\em i.e.}
\[\begin{array}[t]{cccccccc}
  s & =     & a_1 & a_2 & \cdots & a_{2k+1} & a_{2k+2} & \cdots\\
& \lambda_G & \downarrow & \downarrow & & \downarrow & \downarrow\\
    &       & O   & P   &        &   O      &  P \\ 
\end{array} . \]
Thus $P_G$ represents the game tree by the prefix-closed language of 
strings labelling paths from the root. Note that the tree can have 
infinite branches, corresponding to the fact that there can be 
infinite plays in the game. In terms of the representation by 
strings, this would mean that all the finite prefixes of some infinite 
sequence of moves would be valid plays.

For example, the game
$$(\{a_1,a_2,b_1,b_2,b_3\},
\begin{array}[t]{ccccccccccc}
  \{&a_1&,&a_2&,&b_1&,&b_2&,&b_3&\},\\
  &\downarrow&&\downarrow&&\downarrow&&\downarrow&&\downarrow\\
  &O&&O&&P&&P&&P
\end{array}
\{\epsilon,a_1,a_1b_1,a_2,a_2b_2,a_2b_3\})$$
represents the tree
$$
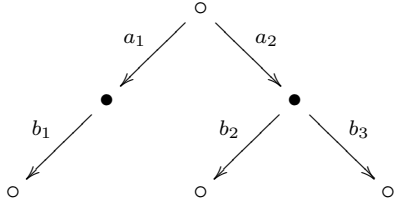
\xymatrix{
  & & {\circ}\ar[dl]_{a_1}\ar[dr]^{a_2}\\
  & {\bullet}\ar[dl]_{b_1}& & {\bullet}\ar[dl]_{b_2}\ar[dr]^{b_3}\\
  {\circ} & & {\circ} & & {\circ}
}$$

We are using games to represent the meaning of \emph{logical formulas} or {\em types}.  A game can be seen as specifying the possible interactions
between a System and its Environment. In the traditional
interpretation of types as structured sets of some kind, types are
used to classify {\em values}. By contrast, games classify {\em
behaviours}. \emph{Proofs} or {\em Programs} will be modelled by {\em strategies}, {\em
i.e.}\ rules specifying how the System should actually play.

Formally, we define a (deterministic) strategy $\sigma$ on a game $G$
to be a non-empty  subset 
$\sigma\subseteq P_G^{\mathrm{even}}$ of the game tree, satisfying:
\[ \begin{array}{ll}
  (\mathbf{s1}) &  \epsilon \in \sigma \\
(\mathbf{s2}) &  sab \in \sigma \ \Longrightarrow\ s \in \sigma \\
(\mathbf{s3}) &   sab,sac\in\sigma\ \Longrightarrow\  b=c . \label{eqn:one}
\end{array} \]

To understand this definition, think of
$$s=a_1 b_1\cdots a_k b_k \in \sigma$$
as a record of repeated interactions with the Environment following
$\sigma$.  It can be read as follows:
\begin{verse}                   
 If the Environment initially does $a_1$,\\*
  \hspace{2em} then respond with $b_1$;\\
  \hspace{2em}If the Environment then does $a_2$,\\*
  \hspace{4em} then respond with $b_2$;\\
  \hspace{10em}$\vdots$\\
  \hspace{8em}If the Environment finally does $a_k$,\\*
  \hspace{10em} then respond with $b_k$.
\end{verse}
The first two conditions on $\sigma$ say that it is a sub-tree of 
$P_{G}$ of even-length paths. The third is a determinacy condition.

This can be seen as generalizing the notion of graph of a relation,
{\em i.e.}\ of a set of ordered pairs, which can be read as
a set of stimulus-response instructions.  The generalization is that
ordinary relations describe a single stimulus-response event only
(giving rules for what the response to any given stimulus may be),
whereas strategies describe repeated interactions between the System
and the Environment.  We can regard $sab\in\sigma$ as saying: `when
given the stimulus $a$ in the context $s$, respond with $b$'.  Note
that, with this reading, the condition (s3) generalizes the
usual single-valuedness condition for (the graphs of) partial
functions.
Thus a useful slogan is:
\begin{quote}
  ``Strategies are (partial) functions extended in time.''
\end{quote}

\begin{example}\rm
Let $\mathbb B$ be the game
\[ (\{ * , \true , \false \} , \{ * \mapsto O , \true \mapsto P, 
\false \mapsto P \} ,
\{\epsilon,*,*\true,*\false\})\]
 
$$\xymatrix{
  & {\circ}\ar[d]^{*}\\
  & {\bullet}\ar[dl]_\true\ar[dr]^\false\\
  {\circ} & & {\circ}
}$$
This game can be seen as representing the data type of booleans.  The
opening move $*$ is a request by Opponent for the data, which can be 
answered by either $\true$ or $\false$ by
Player.  The strategies on $\mathbb B$ are as follows:
$$\begin{array}{ccc}
  \{\epsilon\} & \Pref\{*\true\} & \Pref\{*\false\}
\end{array}$$

The first of these is the undefined strategy (`$\bot$'), the second
and third correspond to the boolean values $\true$ and $\false$.  
Taken
with the inclusion ordering, this ``space of strategies'' corresponds to
the usual flat domain of booleans:
$$
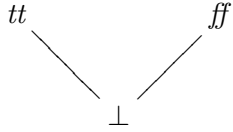
\xymatrix{
  {\true}\ar@{-}[dr] & & {\false}\ar@{-}[dl]\\
                 & \bot 
}$$
\end{example}    

\subsection*{Constructions on games}
We will now describe some fundamental constructions on games.

\subsubsection*{Tensor Product}
Given games $A$, $B$, we describe the tensor product $A\otimes B$.
$$\begin{array}{lcl}
  M_{A\otimes B} & = & M_A + M_B \\
  \lambda_{A\otimes B} & = & [\lambda_A, \lambda_B]\\
  P_{A\otimes B} & = & \{ s\in M_{A\otimes B}^{\mathrm{alt}}\; |\; 
                       s{\restriction} M_A \in P_A 
                         \wedge s{\restriction} M_B \in P_B \}
\end{array}$$

We can think of $A\otimes B$ as allowing play to proceed in {\em both}
the subgames $A$ and $B$ in an interleaved fashion.  It is a form of
`disjoint ({\em i.e.}\ non-communicating or interacting) parallel
composition'.

A first hint of the additional subtleties introduced by the explicit
representation of both System and Environment is given by the
following result.

\begin{proposition} (Switching condition)\\
In any play $s\in P_{A\otimes B}$, if successive moves $s_i$,
$s_{i+1}$ are in different subgames ({\em i.e.}\ one is in $A$ and the
other in $B$), then $\lambda_{A\otimes B}(s_i)=P$, 
$\lambda_{A\otimes B}(s_{i+1})=O$.

In other words, only Opponent can switch from one subgame to another;
Player must always respond in the same subgame that Opponent just
moved in.
\end{proposition}

To prove this, consider for each $s\in P_{A\otimes B}$ the `state'
$$\state{s}=
  (\parity(s\restriction A),\parity(s\restriction B))$$
We will write $O$ for even parity, and $P$ for odd parity, since 
{\em e.g.}\ after a play of even parity, it is Opponent's turn to 
move.
Initially, the state is $\state{\epsilon}=(O,O)$.  Note that O can 
move in
either sub-game in this state.  If O moves in $A$, then the state
changes to $(P,O)$.  P can now only move in the first component.
After he does so, the state is back to $(O,O)$.  Thus we obtain the
following `state transition diagram': 
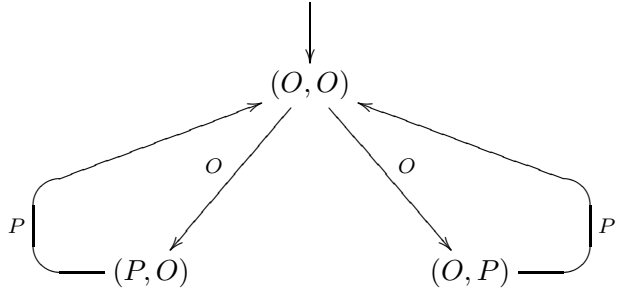
$$\xymatrix{
  & & \ar[d]\\
  & & (O,O)\ar[ddl]_O\ar[ddr]^O\\
  & & & &\\
  & (P,O)\ar`l[ul]`[urr]^P[uur] & & (O,P)\ar`r[ur]`[ull]_P[uul]
}$$

We see immediately from this that the switching condition holds; and
also that the state $(P,P)$ can never be reached ({\em i.e.}\ for no
$s\in P_{A\otimes B}$ is $\state{s}=(P,P)$).

\subsubsection*{Linear Implication}
Given games $A$, $B$, we define the game $A\multimap B$ as follows:
$$\begin{array}{lcl}
  M_{A\multimap B} & = & M_A + M_B \\
  \lambda_{A\otimes B} & = & [\overline{\lambda_A}, \lambda_B]
    \mathrm{\hspace{4em}where}\ {\overline \lambda_A}({\mathit m})=
                         \left\{\begin{array}[c]{l}
                           P\mbox{\rm\ when }\lambda_A(m)=O\\
                           O\mbox{\rm\ when }\lambda_A(m)=P
                        \end{array}\right.\\
  P_{A\multimap B} & = & \{ s\in M_{A\multimap B}^{\mathrm{alt}}\; 
|\; 
                       s\restriction M_A \in P_A 
                         \wedge s\restriction M_B \in P_B \}
\end{array}$$

This definition is {\em almost} the same as that of $A\otimes B$.  The
crucial difference is the inversion of the labelling function on the
moves of $A$, corresponding to the idea that on the left of the arrow
the r\^oles of Player and Opponent are interchanged.

If we think of `function boxes', this is clear enough:
$$
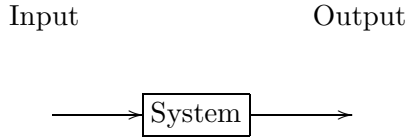
\xymatrix{
  *\txt{Input} & & *\txt{Output}\\ \vspace{-3mm}
  \ar[r] & *+[F]{\mathrm{System}}\ar[r]&
}$$
On the output side, the System is the producer and the Environment is
the consumer; these r\^oles are reversed on the input side.

Note that $M_{A\multimap B}^{\mathrm{alt}}$, and hence 
$P_{A\multimap B}$, are in general quite different to 
$M_{A\otimes B}^{\mathrm{alt}}$, $P_{A\otimes B}$ respectively.  In
particular, the first move in $P_{A\multimap B}$ must always be in
$B$, since the first move must be by Opponent, and all opening moves
in $A$ are labelled $P$ by $\overline{\lambda_A}$.

We obtain the following switching condition for $A\multimap B$:
\begin{quote}
  If two consecutive moves are in different components, the first was
  by Opponent and the second by Player; so only Player can switch
  components.
\end{quote}

This is supported by the following state-transition diagram:
$$
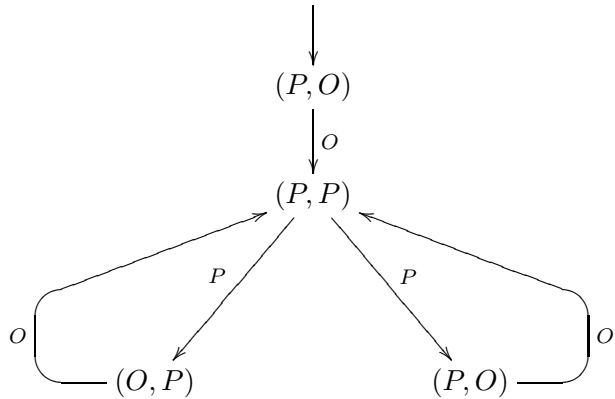
\xymatrix{
  & & \ar[d]\\
  & & (P,O)\ar[d]^{O}\\
  & & (P,P)\ar[ddl]_P\ar[ddr]^P\\
  & & & &\\
  & (O,P)\ar`l[ul]`[urr]^O[uur] & & (P,O)\ar`r[ur]`[ull]_O[uul]
}$$

\begin{example} \rm The copy-cat strategy.

For any game $A$, we define a strategy on $A\multimap A$.  This will
provide the identity morphisms in our category, and the interpretation
of logical axioms $A\vdash A$.

To illustrate this strategy, we undertake by the power of pure logic
to beat a Grand-Master in chess.  To do this, we play two
games, one against, say, Kasparov, as White, and one against
Short)  as
Black.  The situation is as follows:
$$
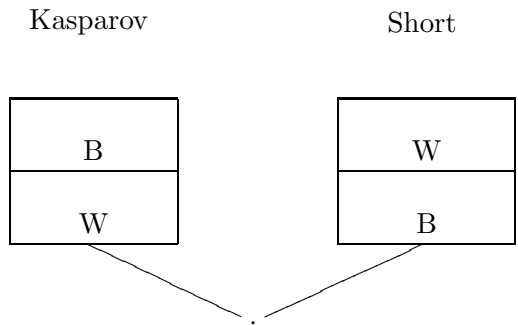
\xymatrix{
  *\txt{Kasparov} & & *\txt{Short}\\
  *\txt{{
    \begin{tabular}{|c|}
      \hline\\
      \mbox{\hspace{2em}B\hspace{2em}}\\
      \hline\\
      W\\
      \hline
    \end{tabular}}}& &
  *\txt{{
    \begin{tabular}{|c|}
      \hline\\
      \mbox{\hspace{2em}W\hspace{2em}}\\
      \hline\\
      B\\
      \hline
    \end{tabular}}} \\
 & {\cdot}\ar@{-}[ul]+D\ar@{-}[ur]+D
}$$

We begin with the game against Short.  He plays his opening move, and
we play his move in our game against Kasparov.  After Kasparov
responds, we play his move as our response to Short.  In this way, we
{\em play the same game twice}, but {\em once as White\/} and {\em
  once as Black}.  Thus, whoever wins, we win one game.  Otherwise
put, we act as a buffer process, indirectly playing Kasparov off
against Short.

This copy-cat process can be seen as a `dynamic tautology', by
contrast with classical propositional tautologies, which are vacuous
static descriptions of states of affairs.  The logical aspect of this
process is a certain `conservation of flow of information' (which
ensures that we win one game).

\begin{exercise}
{\textrm Suppose we had to play in \emph{two games} against Short, both 
  as Black, as well as one game against Kasparov as White.}
$$
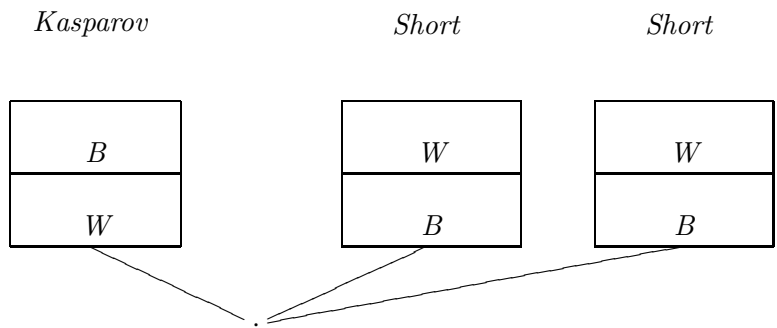
\xymatrix{
  *\txt{Kasparov} & & *\txt{Short} & *\txt{Short} \\
  *\txt{{
    \begin{tabular}{|c|}
      \hline\\
      \mbox{\hspace{2em}B\hspace{2em}}\\
      \hline\\
      W\\
      \hline
    \end{tabular}}}& & 
  *\txt{{
    \begin{tabular}{|c|}
      \hline\\
      \mbox{\hspace{2em}W\hspace{2em}}\\
      \hline\\
      B\\
      \hline
    \end{tabular}}} & 
  *\txt{{
    \begin{tabular}{|c|}
      \hline\\
      \mbox{\hspace{2em}W\hspace{2em}}\\
      \hline\\
      B\\
      \hline
    \end{tabular}}} \\
 & {\cdot}\ar@{-}[ul]+D\ar@{-}[ur]+D\ar@{-}[urr]+D
}$$
{\textrm Would the same idea work?}

{\textrm How about playing in two games against Kasparov, both as White?}
$$
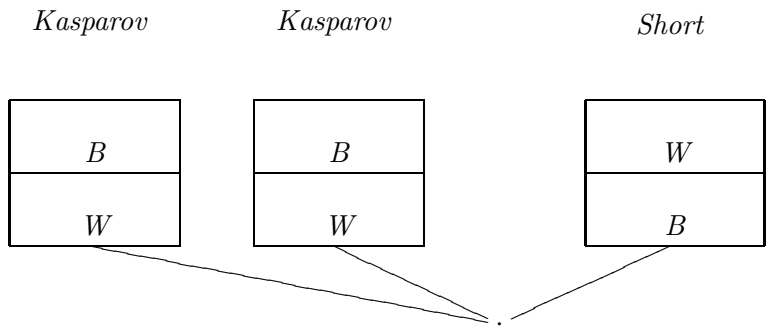
\xymatrix{
  *\txt{Kasparov} & *\txt{Kasparov} &  & *\txt{Short} \\
  *\txt{{
    \begin{tabular}{|c|}
      \hline\\
      \mbox{\hspace{2em}B\hspace{2em}}\\
      \hline\\
      W\\
      \hline
    \end{tabular}}}& 
  *\txt{{
    \begin{tabular}{|c|}
      \hline\\
      \mbox{\hspace{2em}B\hspace{2em}}\\
      \hline\\
      W\\
      \hline
    \end{tabular}}} & &
  *\txt{{
    \begin{tabular}{|c|}
      \hline\\
      \mbox{\hspace{2em}W\hspace{2em}}\\
      \hline\\
      B\\
      \hline
    \end{tabular}}} \\
 & & {\cdot}\ar@{-}[ull]+D\ar@{-}[ul]+D\ar@{-}[ur]+D
}$$
{\textrm Comment on the logical significance of these observations.}
\end{exercise}

In general, a copy-cat strategy on $A$ proceeds as follows:
$$\begin{array}{cccccc}
                   & A   & \multimap & A\\
  \mathrm{Time}\\
  1                &     &           & a_1 & \;\; {\rm O}\\
  2                & a_1 &           &     & \;\; {\rm P}\\
  3                & a_2 &           &     & \;\; {\rm O}\\
  4                &     &           & a_2 & \;\; {\rm P}\\
  \vdots           &     &   \vdots  &     &  \;\;\vdots
\end{array}$$

$$\id{A}=\{s\in P_{A_1\multimap A_2}^{\mathrm{even}} \;\; | \;\;
\forall t \ \mbox{even-length prefix of} \ s: \ 
t{\restriction} A_1 = t{\restriction} A_2\}$$

\noindent (Here, we write $A_1$, $A_2$ to index the two occurrences of
$A$ in $A\multimap A$ for ease of reference.  Note also that we write
$s\restriction A_1$ rather than $s{\restriction} M_{A_1}$.  We will
continue with both these notational ``abuses'').

We indicate such a strategy briefly by 
$\xymatrix@=0pt{A\ar@{-} `u/5pt[r] `/5pt[rr] & \multimap & A}$, 
alluding to
axiom links in the proof nets of Linear Logic.
\end{example}

\begin{example} \rm  Application ({\em Modus Ponens\/}).

$${\mathtt Ap}_{A,B}: (A\multimap B)\otimes A \multimap B$$
This is the conjunction of two copy-cat strategies

$$\xymatrix{
(A\ar@{-} `u/5pt[r] 
            `/5pt[rrrr]
            [rrrr] 
& \multimap & 
B) \ar@{-} `u/10pt[r] 
            `/10pt[rrrr]
            [rrrr] & \otimes&  A& \multimap &  B}$$

Note that $A$ and $B$ each occur once positively and once negatively
in this formula; we simply connect up the positive and negative
occurrences by `copy-cats'.
\begin{center}
${\mathtt Ap}_{A,B}=
    \{s\in P^{\mathrm{even}}_{(A_1\multimap B_1)\otimes A_2\ 
\multimap\  B_2}\; |\ \ \ \ \ \ \ \ \ \ \ \ \ \ \ \ \ \ \ \ \ \ \ \ \ 
\ \ \ \ \ \ \ \ \ \ \ \ \ \ \ \ \ \ \ \ \ \ \ \ \ \ \ \ \ \ \ \ \ \ \ 
\ \ \ \ \ \ \ \ \ \ \ \ \ \ \ \ \ $\\
$\ \ \ \ \ \ \ \ \ \ \ \ \ \ \ \ \ \ \ \ \ \ \ \ \ \ \ \ \ \forall 
t\  \mbox{even-length prefix of} \ s:
\   t{\restriction} A_1 = t{\restriction} A_2\  \wedge\ 
      t{\restriction} B_1 = t{\restriction} B_2 \}$\end{center}

To understand this strategy as a protocol for function application,
consider the following play:\\
\begin{tabular}{ll}
$\begin{array}{cccccccccc}
& ( & A & \multimap & B & ) & \otimes & A & \multimap & B\\
O&&&&&&&&&\mathrm{ro}\\
P&&&&\mathrm{ro}\\
O&&\mathrm{ri}\\
P&&&&&&&\mathrm{ri}\\
O&&&&&&&\mathrm{id}\\
P&&\mathrm{id}\\
O&&&&\mathrm{od}\\
P&&&&&&&&&\mathrm{od}
\end{array}$
&
\begin{tabular}{l}
  ro --- request for output\\
  ri --- request for input\\
  id --- input data\\
  od --- output data
\end{tabular}
\end{tabular}

\vskip 3ex

\noindent
The request for output to the application function is copied to the
output side of the function argument; the function argument's
request for input is copied to the other argument; the input data
provided at the second argument is copied back to the function
argument; the output from the function argument is copied back to
answer the original request. It is a protocol for {\em linear} 
function
application since the state of both the function and the argument will
change as we interact with them; we have no way of returning to the
original state. Thus we ``consume'' our ``resources'' as we produce
the output. In this way there is a natural match between game
semantics and linear logic.
\end{example}

\subsection*{The Category of Games ${\cal G}$}

\begin{itemize}
\item Objects: Games

\item Morphisms: $\sigma : A \RED B$ are strategies $\sigma$ on 
               $A \multimap B$.

\item Composition: interaction between strategies.
\end{itemize}

This interaction can be described as ``parallel composition plus hiding''.

\[
\frac
{\sigma : A \rightarrow B \quad \tau: B \rightarrow C}
{\sigma ;\tau: A \rightarrow C}
\]
$\sigma ;\tau 
\ = \ (\sigma \PARA \tau )/ B 
\ = \ \ASET{s{\RES} A,C \ | \ s\in \sigma \PARA \tau }$\\[2mm]
$\sigma \PARA \tau 
\ = \ \ASET{s\in (M_A + M_B + M_C)^\ast \ | \ 
   s{\RES} A,B \in {\sigma} 
    \ \wedge \  
    s{\RES} B,C \in {\tau}}$.\\[2mm]

(Note that we extend our abuse of notation for restriction here; by 
$s {\RES} A,B$ we mean the restriction of $s$ to $M_{A} + 
M_{B}$ as a ``subset'' of $M_{A} + M_{B} + M_{C}$, and similarly for 
$s{\RES}A,C$ and $s{\RES}B,C$.)
This definition looks very symmetric, but the actual possibilities are
highly constrained by the switching condition. 

$$\begin{array}{cccccccc}
&A & \stackrel{\sigma}{\multimap} & B & \;\;\; &B & \stackrel{\tau}{\multimap} & C\\
&  & &     & &     & & c_1\\
&  & &     & & b_1 & & \\
&  & & b_1 & &     & & \\
&  & & b_2 & &     & & \\
&  & &     & & b_2 & & \\
&  & & \vdots & & \vdots & & \\
&  & &     & & b_k & & \\
&  & & b_k & &     & & \\
&a_1\\
\end{array}$$

\noindent Initially, Opponent must move in $C$ (say with $c_1$).
We consider $\tau$'s response. If this is in $C$, then this is the 
response of \ $\sigma ; \tau$ \ to \ $c_1$. If $\tau$ responds in $B$, 
say with $b_1$, then a move by Player in $B$ in $B\multimap C$ is
a move by Opponent in 
$A \multimap B$. So it makes sense to consider $\sigma$'s response to 
$b_1$. If it is in $A$, this is the overall response of 
$\sigma ; \tau$ to $c_1$. If $\sigma$ responds with 
$b_2$ in $B$, then $b_2$ is a move by Opponent in 
$B \multimap C$, and we consider $\tau$'s response. Continuing 
in this way, we obtain a uniquely determined sequence. 
\[
c_1 b_1 b_2 \cdots b_k \cdots
\]
If the sequence ends in a visible action in $A$ or $C$, 
this is the response by the strategy $\sigma ; \tau$ to the 
initial move $c_1$, with the internal dialogue between $\sigma$ and 
$\tau$ in $B$ being hidden from the Environment. 
Note that $\sigma$ and $\tau$ may continue their internal dialogue 
in $B$ forever. This is ``infinite chattering'' in CSP
terminology, and ``divergence by an infinite $\tau$-computation'' 
in CCS terminology. 

As this discussion clearly shows composition in ${\cal G}$ 
is interaction between strategies. 
The following fact is useful in the analysis of composition. 

The map \ $s \mapsto s{\RES} A, C$ induces a surjective map
\[ 
\psi: \sigma \PARA \tau \ \RED \ \sigma ; \tau
\]

\noindent {\bf Covering Lemma.} \ \ $\psi$ is injective 
(and hence bijective) so for each $t \in \sigma ; \tau$ there is 
a unique $s\in \sigma \PARA \tau$ such that 
$s\RES A, C \ = t$. 

\begin{quote}
If $t = m_1m_2 .... m_k$, 
then $s$ has the form 
\[ m_1 u_1 m_2 u_2  .... u_{k-1}m_k \]
where $u_i \in M_B^\ast$, $1 \leq i < k$. 
\end{quote}

\begin{exercise} \rm 
Prove the Covering lemma by formalizing the preceding 
discussion. 
\end{exercise}

\subsubsection*{An alternative definition of Cut}

We give a more direct, `computational' definition.

\[ \sigma ; \tau = \{ s;t \mid s \in \sigma \wedge t \in \tau \wedge s
\proj B = t \proj B \} . \]
This defines Cut `pointwise' via an operation on single plays. This
latter operation is defined by mutual recursion of four operations
covering the following 
situations:
\begin{center}
\begin{tabular}{lll}
1. & $s \lmerge t$ & O is to move in $A$.\\
2. & $s \rmerge t$ & O is to move in $C$. \\
3. & $s \bbslash t$ & $\sigma$ to move. \\
4. & $s \sslash t$ & $\tau$ to move. \\
\end{tabular}
\end{center}

\[ \begin{array}{cccr}
\gamma s \lmerge t & = & \gamma (s \bbslash t) & \\
\ve \lmerge t & = & \ve & \\
s \rmerge bt & = & b(s \sslash t) & \\
s \rmerge \ve & = & \ve & \\
\gamma s \bbslash t & = & \gamma (s \lmerge t) & (\gamma \in
M_{\Gamma} ) \\
as \bbslash at & = & s \sslash t  & (a \in M_A ) \\
s \sslash bt & = & b (s \rmerge t) & (b \in M_B ) \\
as \sslash at & = & s \bbslash t & (a \in M_A )
\end{array} \]
We can then define
\[ s;t \; = \; s \rmerge t . \]

\begin{exercise} \rm 
Prove that the two definitions of $\sigma ; \tau$ coincide. 
\end{exercise}

\begin{proposition}
${\cal G}$ is a category.
\end{proposition}

In particular, ${\tt id}_A: A \RED A$ is the copy-cat strategy 
described previously. 

\begin{exercise} \rm
Verify this Proposition.
\end{exercise}

\begin{exercise} \rm
Define a strategy \ 
{\tt not}: ${\mathbb B} \RED {\Bbb B}$ 
\ on the boolean game defined previously to 
represent Boolean complement. Calculate explicitly the strategies 
\[ 
\bot; \mbox{{\tt not}} \quad \quad 
\true; \mbox{{\tt not}} \quad \quad 
\false; \mbox{{\tt not}} 
\]
and hence show that this strategy does indeed represent the intended function.
(For this purpose, treat strategies
$\sigma$ on ${\Bbb B}$ as strategies 
$\sigma: I \RED {\Bbb B}$ where 
\[ I \ = \ (\emptyset,\ \emptyset, \ \ASET{\varepsilon})\] 
is the empty game, so that the above compositions make sense). 
\end{exercise}

\begin{exercise} \rm
Embed the category of sets and partial functions faithfully into ${\cal G}$.
Is your embedding full?
What about the category of flat domains and monotone maps?
\end{exercise}

\subsection*{Tensor structure of ${\cal G}$}

We will now see (in outline) that ${\cal G}$ is an ``autonomous'' 
$=$ symmetric monoidal closed category, and hence a model for IMLL,
Intuitionistic Multiplicative Linear Logic. 

We have already defined the tensor product $A \otimes B$ on objects. 
Now we extend it to morphisms:
\[
\frac
{\sigma :  A \rightarrow B \quad \tau: A' \rightarrow B'}
{\sigma \otimes \tau: A \otimes A' \rightarrow B\otimes B'}
\]
$\sigma \otimes \tau\ \ = \ \ 
\{ s\in P_{A \otimes A'\multimap B\otimes B'}^{\mbox{\scriptsize even}}
\ \ | \ \ s\RES A, B \in \sigma \ \ \wedge\  \ s\RES A',B' \in \tau\}$.\\

\noindent This can be seen as disjoint (i.e. non-communicating) parallel 
composition of $\sigma$ and $\tau$. \\

\begin{exercise} \rm
Check functoriality, i.e. the equations 
\begin{itemize}
\item $(\sigma \otimes \tau ) ; (\sigma' \otimes \tau' )
\ = \ (\sigma ; \sigma' ) \otimes (\tau ; \tau' )$.
\item ${\tt id}_{A} \otimes {\tt id}_{B} 
\  = \ {\tt id}_{A\otimes B}$.
\end{itemize}
The tensor unit is defined by:
\[ 
I \ = \ (\emptyset, \ \emptyset, \ \ASET{\varepsilon})
\]
\end{exercise}

The canonical isomorphisms are conjunctions of copy-cat strategies.\\
\[
\begin{array}{rc}
{\mathtt assoc}_{A,B,C} : 
& (A \otimes B)\otimes C \SIMRED A \otimes (B\otimes C)\\[5mm]

&
{\xymatrix@=5pt@!
{ (A \ar@{-} `u/4pt[r] 
            `/4pt[rrrrrr]
            [rrrrrr] 
& \otimes &  
  B) \ar@{-} `u/8pt[r] 
            `/8pt[rrrrrr]
            [rrrrrr] 
& \otimes &  
  C \ar@{-} `u/12pt[r] 
            `/12pt[rrrrrr]
            [rrrrrr]  
            & \multimap &
A & \otimes  & (B & \otimes  & C) & }}\\[8mm]

{\mathtt symm}_{A,B} : 
& A \otimes B \SIMRED B \otimes A\\[5mm]
&{\xymatrix@=5pt@!
{ A \ar@{-} `u/10pt[r] 
            `/10pt[rrrrrr]
            [rrrrrr]
& \otimes&  
B \ar@{-} `u/5pt[r] 
            `/5pt[rr]
            [rr]
  & \multimap & B & \otimes & A }}\\[8mm]

{\mathtt unitl}_{A} : 
& (I \otimes A) \SIMRED A \\[5mm]
&{\xymatrix@=5pt@!{
(I & \otimes & 
A\ar@{-} `u/5pt[r] 
            `/5pt[rr]
            [rr]
) & \multimap &  A}}\\[8mm]

{\mathtt unitr}_{A} : 
&(A \otimes I) \SIMRED A \\[5mm]
&{\xymatrix@=5pt@!{
(A \ar@{-} `u/5pt[r] 
            `/5pt[rrrr]
            [rrrr] 
& \otimes &  I) & \multimap & A}}\\
\end{array}
\]


The application (or evaluation) morphisms
\[ 
{\mathtt Ap}_{A,B}: (A\multimap B)\otimes A \RED B
\]
have already been defined. For currying, given 
\[ 
\sigma : A \otimes B \multimap C
\]
define 
\[
\Lambda(\sigma): A \RED (B \multimap C)
\]
by 
\[
\Lambda(\sigma) = \ASET{\alpha^\ast(s)
\ |\ 
s\in\sigma}
\]
where $\alpha: 
(M_A + M_B) + M_C \SIMRED 
M_A + (M_B + M_C)$
 is the canonical isomorphism in $\Set$. \\

\begin{exercise}\rm
Verify that the above definitions work!
{\em E.g.} verify the equations  ${\mathtt Ap}\circ (\Lambda(\sigma)\otimes 
\mathrm{id}_{A}) \ = \ \sigma$:
%
%
\begin{diagram}
(A \multimap B)\otimes A & \rTo^{\mathtt Ap}  & B \\
\uTo^{\Lambda(\sigma)\otimes{\id{A}}} & \ruTo_{\sigma} \\
C\otimes A
\end{diagram}
and $\Lambda({\mathtt Ap}\circ (\tau\otimes \id{A}))
\ = 
\ \tau$ 
\ for 
\ $\tau : C \RED (A\multimap B)$. 
\end{exercise}

\begin{exercise} \rm
Prove that $I$ is terminal in 
${\cal G}$, i.e. for each $A$ there is a unique morphism 
$t_A  : A \RED I$. 
\end{exercise}

This shows that ${\cal G}$ is really a model of Affine Logic, 
in which (unlike in Linear Logic proper) the Weakening rule is valid.
Indeed, tensor has ``projections'':
\[ 
A\otimes B  
\ {\stackrel{\mathrm{id}_{A}\otimes t_B}{\RED}}
\ A\otimes I
\ \stackrel{\mathrm{unitr}}{\SIMRED}
\ A.
\] 

\begin{exercise} \rm
Given $A,B$ define $A \& B$ by 
$$\begin{array}{lcl}
  M_{A\& B} & = & M_A + M_B \\
  \lambda_{A\& B} & = & [\lambda_A, \lambda_B]\\
  P_{A\& B} & = & \{ \mathtt{inl}^{\ast}(s) \mid s \in P_{A} \} \cup
  \{ \mathtt{inr}^{\ast}(t) \mid t \in P_{B} \} .
\end{array}$$
(Draw a picture of the game tree of $A\& B$; it is formed by gluing 
together the trees for $A$ and $B$ at the root. There is no overlap 
because we take the disjoint union of the alphabets.)
Prove that $A\& B$ is the product of $A$ and $B$ in ${\cal G}$, 
i.e. define projections
\[ 
A 
\ \stackrel{{\mathtt fst}}{\longleftarrow}
\ A\& B 
\ \stackrel{{\mathtt snd}}{\RED}
\ B
\]
and pairing 
\[ 
\ENCan{\ , \ }: 
{\cal G}(C,A) \times {\cal G}(C,B)
\RED 
{\cal G}(C,A\& B) 
\]
and verify the equations 
$$\begin{array}{lcl}
  \ENCan{\sigma,\:\tau} ; {\mathtt fst} & = & \sigma\\
  \ENCan{\sigma,\:\tau} ; {\mathtt snd} & = & \tau\\
  \ENCan{v ; {\mathtt fst}, \: v ; {\mathtt snd}} & = & v 
\quad {\mathrm{for}} \ 
v: C \RED A \& B\\
\end{array}$$
\end{exercise}

\begin{exercise} \rm
Try to define coproducts in ${\cal G}$. What is the problem?
\end{exercise}

\begin{exercise} \rm
A strategy $\sigma$ on $A$ is {\em history-free} if it satisfies
\begin{itemize}
\item $sab,\ tac \ \in \ \sigma \THEN b = c$.
\item $sab,\ t \ \in \ \sigma, \ ta\in P_A\THEN tab \in \sigma$.
\end{itemize}
Prove that $\id{A}$, ${\mathtt assoc}_{A,B,C}$, 
${\mathtt sym}_{A,B}$, ${\mathtt Ap}_{A,B}$, 
${\mathtt unitl}_{A}$, ${\mathtt unitr}_{A}$, 
${\mathtt fst}_{A,B}$, ${\mathtt snd}_{A,B}$ 
are all history-free; and that if $\sigma$ and $\tau$ are history free so are 
$\sigma \: ; \: \tau$, $\sigma \otimes \tau$, and 
$\Lambda(\sigma)$. 
Conclude that the sub-category ${\cal G}^{\mbox{\scriptsize hf}}$, 
of history-free strategies is also a model of IMLL. 
What about the pairing operation 
$\ENCan{\sigma,\:\tau}$? Does ${\cal G}^{\mbox{\scriptsize hf}}$ 
have binary products?
\end{exercise}

\renewcommand{\rel}[1]{{\cal{R}}_{#1}}
\renewcommand{\Rel}{\mbox{\bf Rel}}
\renewcommand{\Set}{\mbox{\bf Set}}
\renewcommand{\Span}{\mbox{\bf Span}}


\section{Winning Strategies}
As we have seen, deterministic strategies can be viewed as partial 
functions
extended in time. This partiality is appropriate when we aim to model 
programming languages with general recursion, in which the 
possibility of non-termination arises. However we would also like to use game semantics 
to model
logical systems satisfying Cut Elimination or Strong Normalization. 
We would 
therefore like to find a condition on strategies generalizing 
totality of 
functions. The obvious candidate is to require that at each stage of 
play, a 
strategy $\sigma$ on A has some response to every possible move by 
opponent.

\[ {\mathbf(tot)} \hspace{3em}   s \in \sigma,      sa\in\P{A}
      \Rightarrow     \exists b:\ sab \in \sigma \]

\noindent
Call a strategy \emph{total} if it satisfies this condition. 
However, totality as so defined does not suffice ; 
in particular, it is not closed under composition.

\begin{exercise} \rm
Find games $A, B, C$ and strategies $\sigma:A\rightarrow B$ 
and $\tau:B\rightarrow C$, such that
\begin{itemize} 
\item $\sigma$ and $\tau$ are total 
\item $\sigma;\tau$ is not total.
\end{itemize}
(Hint: use infinite chattering in $B$.)
\end{exercise}
 
The best analogy for  understanding this fact is with the untyped 
$\lambda$-calculus: the class of strongly normalizing terms is not 
closed under
application. Thus in the  Tait/Girard method for proving strong 
normalization 
in various systems of typed $\lambda$-calculus, one introduces a 
stronger 
property which does ensure  closure under application. 
The approach we will pursue with strategies can be seen 
as a semantic analogue of this idea.

The idea is to take \emph{winning} strategies. 
Given a game A, define $\P{A}^{\infty}$, the infinite plays over A, 
by 
\[ \P{A}^{\infty} = \{s \in M_{A}^{\omega} \; | \; \Pref(s) \subseteq 
\P{A} \} \]
(By $\Pref(s)$ we mean the set of \emph{finite} prefixes.)
Thus the infinite plays correspond exactly to the infinite 
branches of the 
game tree.

Now  a set $W \subseteq \P{A}^{\infty}$
can be 
interpreted as designating those infinite plays which are ``wins'' 
for Player.
We say that $\sigma$ is a \emph{winning strategy} with respect to W 
(notation: $\sigma \models W$), if: 
\begin{itemize}
\item $\sigma$ is total
\item \{s $\in \P{A}^{\infty} \ | \ \Pref(s) \subseteq \sigma \} 
\subseteq W$.
\end{itemize}
Thus $\sigma$ is winning if at each finite stage when it is Player's 
turn to move it has a well defined response, and moreover every 
infinite play following $\sigma$ is a win for Player.

We introduce an expanded of refined notion of game as a pair $(A,
W_{A})$, where $A$ is a game as before, and $W_{A} \subseteq \P{A}^{\infty}$ 
is the designated set of winning infinite plays for Player. 
A winnining strategy for $(A,
W_{A})$ is a strategy for $A$ which is winning with respect to $W_A$.

We now extend the definitions of $\otimes$ and $\linimpl$ to act on
the winning set specifications:
\[ \begin{array}{lcl}
(A, W_{A}) \otimes (B, W_{B}) & = & (A \otimes B, W_{A\otimes B}) \\
(A, W_{A}) \linimpl (B, W_{B}) & = & (A \linimpl B, W_{A\linimpl B}) 
\end{array} \]
where
\[ \begin{array}{lcl}
W_{A\otimes B} & = & \{s \in \P{A \otimes B }^{\infty} \ | \
s\upharpoonright A \in \P{A} \cup W_A
\wedge s\upharpoonright B \in \P{B} \cup W_B \}\\
W_{A\linimpl B} & = & \{s \in \P{A \linimpl B }^{\infty} \ | \ 
s\upharpoonright A \in \P{A} \cup W_A \Rightarrow s \upharpoonright B 
\in W_B \}
\end{array} \]

\begin{exercise}
\rm Why did we not define
\[ W_{A\otimes B} \; = \; \{s \in \P{A \otimes B }^{\infty} \ | \
s\upharpoonright A \in W_{A} 
\wedge s\upharpoonright B \in W_{B} \} ? \]
(Hint: consider the switching condition for $\otimes$).
\end{exercise}
\noindent
In order to check that these definitions work well, we must show that
the constructions on strategies we have introduced in order th model
the proof rules of Linear Logic are well-defined with respect to
winning strategies.

\begin{exercise}
\rm Show that, for any $(A, W_A )$, the copy-cat strategy $\id{A}$ is a
winning strategy.
\end{exercise}

\noindent Now we consider the crucial case of the Cut rule.

Suppose then that $\sigma:(A,W_{A}) \linimpl (B,W_{B})$ and 
$\tau:(B,W_{B}) \linimpl (C,W_{C})$.
We want to prove that $\sigma;\tau$ is total, i.e. that there can be 
no infinite chattering in B. 
\\ Suppose for a contradiction that there is an infinite play 
$$ t = s b_{0}b_{1}\cdots \in  \sigma \| \tau$$
with all moves after the finite prefix $\mathit{s}$ in $B$. 
Then $t\upharpoonright A,B$ 
is an infinite play in $A\multimap B$ following $\sigma$, while 
$t\upharpoonright B,C$
is an infinite play in $B\multimap C$ following $\tau$. 
Since $\sigma$ is winning and $t\upharpoonright A$ is finite, 
we must have $t\upharpoonright B \in W_{B}$. 
But then since $\tau$ is winning we must have $t\upharpoonright C \in 
W_{C}$, 
which is impossible since $t\upharpoonright C$ is finite.

\begin{exercise} 
\rm Give a direct proof (not using proof by 
contradiction) that winning stratregies compose.
\end{exercise}

\begin{exercise}
\textrm{Prove that $\id{A}$, ${\mathtt assoc}_{A,B,C}$, 
${\mathtt sym}_{A,B}$, ${\mathtt Ap}_{A,B}$, 
${\mathtt unitl}_{A}$, ${\mathtt unitr}_{A}$, 
${\mathtt fst}_{A,B}$, ${\mathtt snd}_{A,B}$ 
are all winning strategies; and that if $\sigma$ and $\tau$ are
winning, so are 
$\sigma \: ; \: \tau$, $\sigma \otimes \tau$, $\langle \sigma,\:\tau
\rangle$, and 
$\Lambda(\sigma)$. }
\end{exercise}

\begin{exercise} \rm
Verify that the total strategies 
$$\sigma:\mathbb{B} \rightarrow \mathbb{B}$$ 
correspond exactly to the total functions on the booleans.
\end{exercise}

\begin{exercise} \rm
Consider a game of binary streams \emph{Str}
$$
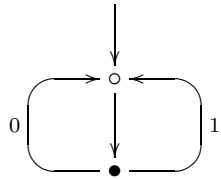
\xymatrix{
  & & \ar[d] & &\\
  & & \circ\ar[d] & &\\
  & & \bullet\ar`l[ul]`[urr]^0[u]\ar`r[ur]`[ull]_1[u] & &
}$$
with plays $*b_{1}*b_{2}*b_{3}\dots$, alternating between requests for
data by Opponent and bits supplied by Player. Let $W_{Str}$ be all
infinite plays of this game.

\noindent
Verify that the winning strategies on ($Str$, $W_{Str}$) correspond 
exactly to the infinite binary sequences. Verify that the winning 
strategies 
$$\sigma:(Str,W_{Str}) \rightarrow (Str,W_{str})$$ 
induce functions which map infinite streams to infinite streams. 
Can you characterize exactly which functions on the domain 
$$\{0,1\}^{*} \cup \{0,1\}^{\omega}$$
with the prefix ordering are induced by winning strategies?
\end{exercise}

\section{Polymorphism}
Our aim now is to use game semantics to give a model
for polymorphism. We extend our notation for types with type variables 
$X,Y,...$ and with second order quantifiers
$$\forall X.A$$

As a test case, we want our model to have the property that the interpretation
it yields of the polymorphic (affine) booleans
$$\forall X.X \multimap (X\multimap X)$$
has only two elements, corresponding to the denotations of the terms
$$\true\eqdef \Lambda X.\lambda x,y:X.x$$
$$\false\eqdef \Lambda X.\lambda x,y:X.y$$
Firstly, we need some control over the {\em size} of the universe of types.
To achieve this, we assume a non empty set $\V$ satisfying
$$\V+\V \subseteq \V$$
(for example take $\V=\{0,1\}^{*}$).

\noindent
Now we define a game $\U$ by:
$$\M{\U} =\V+\V$$
$$\lam{\U}=[{\mathbf K}P,{\mathbf K}O]$$
$$\P{\U}=\M{\U}^{alt} .$$
(Here ${\mathbf K}P$ is the constant function valued at $P$.)
We can define a partial order on games by:
$$A\po B\ \ \eqdef\ \ \M{A}\subseteq\M{B} \ \wedge \ 
\lam{A}=\lam{B}\restriction\M{A} \ \wedge \ \P{A}\subseteq\P{B}$$
Now define $$\G_{\U}=\{ A\in\Obj{\G} \ | \ A\po \U\}$$

We define a {\em variable type} (in $k$ variables)
to be a function (monotone with respect to $\po$)
$$F:\G_\U^{k} \rightarrow \G_{\U}$$
Note that
$$A,B\in\G_{\U} \ \Rightarrow \ A\tensor B, A\multimap B\in\G_{\U}$$
(that was the point of having $\V+\V\subseteq \V$)

\begin{exercise} \rm (If you care about details)
The above is not {\em quite} true. Amend the definition of $A\tensor B$,
$A\multimap B$ slightly to make it true.
\end{exercise}

Thus variable types will be closed under $\tensor$ and $\multimap$.
Given $F,G: \G_{\U}^k \rightarrow \G_{\U}$, we can define
$$F\tensor G(\vec{A}) \ \ = \ \ F(\vec{A}) \tensor G(\vec{A})$$
$$F\multimap G(\vec{A}) \ \ = \ \ F(\vec{A}) \multimap G(\vec{A})$$

A \emph{uniform strategy} $\sigma$ on a variable type $F$ is defined 
to be a strategy on $F(\vec{\U})$ such that, 
for all $\vec{A}\in\G_{\U}^k$, $\sigma_{\vec{A}}$ 
is a well-defined strategy on $F(\vec{A})$, 
where $\sigma_{\vec{A}}$ is defined inductively by
$$\sigma_{\vec{A}}= \{\epsilon\} \ \cup \ \{sab \ | \ s\in\sigma_{\vec{A}},
sa\in\P{F(\vec{A})}, sab \in \sigma\}$$
(NB: in this notation, $\sigma=\sigma_{\vec{\U}}$).

\begin{exercise}\rm
Show that the following properties hold for a uniform strategy $\sigma$ on $F$:

\begin{tabular}{cl}
(i)  &
$\vec{A}\po\vec{B}\ \mbox{(component-wise)}
\Rightarrow \sigma_{\vec{A}}
=\sigma_{\vec{B}}\cap P_{F(\vec{A})}\subseteq \sigma_{\vec{B}}$\\
(ii) & if $(\vec{A}_i\ | i\in \I)$ is a $\po$-directed family in $\G_{\U}^k$,
then \\
& $\sigma_{\veedirect{i\in\I}{\vec{A}_i}}
=\cupdirect{i\in\I}{\sigma_{\vec{A}_i}}$ $\ \ \ \ $ where\\
& $\veedirect{i\in\I}{\vec{A}_i}$ is the directed join
of the $\vec{A}_i$ (defined by component-wise union),\\
& and $\cupdirect{i\in\I}{\sigma_{\vec{A}_i}}$ is the directed union 
of the strategies $\sigma_{\vec{A}_i}$.
\end{tabular}
\end{exercise}

\vspace{2em}
\noindent
Our aim now is to show that, for each $k\in\omega$,
we obtain a category $\G(k)$ with:

\begin{tabular}{lll}
objects & : & variable types  $F:\G_{\U}^k \rightarrow \G_{\U}$\\
morphisms & : & $\sigma:F\rightarrow G$ are uniform strategies $\sigma$ 
on $F\multimap G$
\end{tabular}

\noindent
Moreover $\G(k)$ is an autonomous category.

\vspace{1em}
\noindent
The idea is that all the structure is transferred pointwise 
from $\G$ to $\G(k)$.
E.g if $\sigma:F\multimap G$, $\tau:G\multimap H$, then
$\sigma ; \tau: F\rightarrow H\ \mbox{is given by}\ 
\sigma; \tau = \sigma_{\vec{\U}}; \tau_{\vec{\U}}$.

\begin{exercise}\rm
Check that $\sigma;\tau$ is a well-defined uniform strategy on $F\multimap H$.
\end{exercise}

Similarly, we define 
$$\id{F}= \id{F(\vec{\U})}$$
$$\Ap{F}{G}= \Ap{F(\vec{\U})}{G(\vec{\U})}$$
etc.

Now we define a ``base category'' $\B$ 
with the objects $\G_{\U}^k$, $k\in\omega$,
and $\po$-monotone functions as morphisms.
For each object $\G_{\U}^{k}$ of $\B$, we have the autonomous category
$\G(k)$. 
For each monotone
$$F=\langle F_1, \dots, F_l \rangle: \G_{\U}^{k}\rightarrow
\G_{\U}^{l}$$
we can define a functor
$$F^*:\G(l)\rightarrow \G(k)$$
by 
$$F^*(G)(\vec{A})= G(F(\vec{A}))$$
$$F^{*}(\sigma_{\vec{A}})= \sigma_{F(\vec{A})}$$

\begin{proposition}
The above defines a (strict) indexed autonomous category.
\end{proposition}

\vspace{1em}
\noindent
At this point, we have enough structure to interpret types and terms 
with type variables. It remains to interpret the quantifiers.
For notational simplicity, we shall focus on the case 
$\forall X.A(X)$ where $X$ is the only type variable free in $A$.
Semantically A will be interpreted by a function 
$F:\G_{\U}\rightarrow\G_{\U}$.
We must define a game $\Pi(F)\in\G_{\U}$ as the interpretation of $\forall X.A$

Corresponding to the polymorphic type inference rule
\begin{tabular}{cc}
${(\forall-\mbox{elim})}$ &
$\frac{
\Gamma\vdash t:\forall X.A}
{\Gamma\vdash t\{B\}: A[B/X]}$
\end{tabular}
we  must define a uniform strategy
$$\pi: \K\Pi(F) \rightarrow F .$$
(Here $\K\Pi(F):\G_{\U}\rightarrow\G_{\U}$ is the constant function valued at 
$\Pi(F)$.
Note that $\K = t^{*}_{\U}$ where $t:\U\rightarrow\one=\G_{\U}^{0}$
is the map to the terminal object in $\B$.)

Corresponding to the type inference rule
\[ (\forall - \mbox{intro}) \quad
\frac{\Gamma\vdash t:A}{\Gamma\vdash \Lambda X.t:\forall X.A} \quad  
\mbox{if} \;\; X\not\in
\mbox{FTV}(\Gamma )
\]
we must prove the following universal property:
\begin{quote}
for every $C\in\G_{\U}$ and uniform strategy $\sigma:\K C\rightarrow F$
there exists a unique strategy 
$\Lambda^2(\sigma) : C\rightarrow\Pi(F)$ such that 
\end{quote}
\begin{diagram}
\K\Pi(F)                  & \rTo^{\pi}    & F \\
\uTo^{\K\Lambda^2(\sigma)}& \ruTo_{\sigma} &\\
\K C& & \\
\end{diagram}

This says that there is an adjunction
\begin{diagram}
\G_{\U}=\G_{\U}(0) & \pile{\rTo^{t_{\U}^*}\\ 
\perp\\ \lTo_{\Pi(F)}} & \G_{\U}(1)\\
\end{diagram}

Furthermore, we must show that the Beck-Chevalley condition holds 
(see (Crole 1994)).

\begin{remark}
More generally, we should show the existence of adjunctions
\begin{diagram}
\G_{\U}=\G_{\U}(k) & \pile{\rTo^{p^*}\\ \perp\\ \lTo_{\Pi_{k}(F)}}
&\G_{\U}(k+1)\\
\end{diagram}
where $p:\G^{k+1}_{\U}\rightarrow \G_{\U}^{k}$ is the projection function.
\end{remark}

Now, how are we to construct the game $\Pi(F)$?
Logically, $\Pi$ is a second-order quantifier.
Player must undertake to defend $F$ at any
instance $F(A)$, where $A$ is specified by Opponent.
If Opponent were to specify the entire instance $A$ at the start of the game,
this would in general require an infinite amount of information
to be specified in a finite time, violating a basic continuity principle 
of computation (``Scott's axiom'').
Instead we propose the metaphor of the ``veil of ignorance''
({\sl cf.} {\rm John Rawls}, {\sl A Theory of Justice}).
That is, initially nothing is known about which instance we are playing in.
Opponent progressively reveals the ``game board'' ; 
at each stage, Player is constrained to play within the instance
{\em thus far revealed} by Opponent.

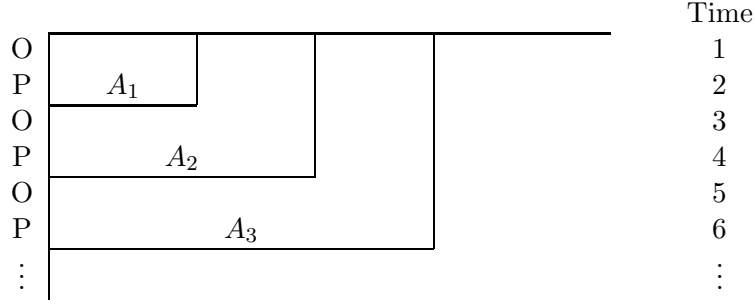
\begin{center}
\begin{tabular}{c|c|c|c|ccc}
\multicolumn{5}{c}{}& & Time\\ \cline{2-5}
O & \hspace{4em} & \hspace{3em} & \hspace{3em} & \hspace{5em} & \hspace{1em} & 1\\
P &  $A_1$ & & & & & 2\\ \cline{2-2}
O & \multicolumn{2}{c|}{} & & & & 3\\
P & \multicolumn{2}{c|}{$A_2$} & & & & 4\\ \cline{2-3}
O & \multicolumn{3}{c|}{}  & & & 5\\
P & \multicolumn{3}{c|}{$A_3$}  & & & 6\\ \cline{2-4}
$\vdots$ & \multicolumn{3}{c}{} & & & $\vdots$\\
\end{tabular}
\end{center}

\noindent
This intuition is captured by the following definition.
$$\M{\Pi(F)} =  \M{F(\U)}$$
$$\lam{\Pi(F)} = \lam{F(\U)}$$
$\P{\Pi(F)}$ is defined inductively as follows:

$\begin{array}{llll}
\P{\Pi(F)}&=& &\{\epsilon\} \\
 & & \cup & \{sa \ |\ s\in\P{\Pi(F)}^{\even} \ \wedge\ \exists A. 
sa\in\P{F(A)}\}\\
        & & \cup & \{sab \ | \ sa\in\P{\Pi(F)}^{\odd} \  
\wedge \ \forall A. sa\in\P{F(A)}\ \Rightarrow \ sab\in\P{F(A)}\}
\end{array}$

\vspace{1em}
The first clause in the definition of $P_{\Pi(F)}$
is the basis of the induction.
The second clause refers to positions in which it is Opponent's turn to move.
It says that Opponent may play in any way which is valid in {\em some}
instance (extending the current one). The final clause refers
to positions in which it is Player's turn to move.
It says that Player can only move in a fashion which is valid in {\em every}
possible instance.

For the polymorphic projection
$$\Pi(F)\stackrel{\pi_A}{\rightarrow}F(A)$$
$\pi_{A}$ plays copy-cat between $\Pi(F)$ and $F(A)$.
This is uniform, witnessed by the ``global copy-cat'' $\id{F(\U)}$.

Why does this definition work?
Consider the situation

$$
\begin{array}{ccc}
\Pi(F) & \rightarrow & F(A) \\
&&a \\
a\\
\end{array}
$$
At this stage, it is Opponent's turn to move,
and of course there are many moves in $\Pi(F)$
which would not be valid in $F(A)$.
However, Opponent in $\Pi(F)$ in contravariant (i.e negative)
position must play as Player in $\Pi(F)$,
and hence is constrained to respond to $a$ only in a fashion
which is valid in {\em every} instance
in which $a$ can be played, and which in particular is valid in $F(A)$.
Hence Opponent's response can safely be copied back into $F(A)$.

\vspace{1em}
\noindent
Now for the universal property.
Given uniform $\sigma:\K C\rightarrow F$, we define 
$$\Lambda^2(\sigma)=\sigma:C\rightarrow\Pi(F)$$
That this is valid follows from the uniformity of $\sigma$ so that at each
stage its response must be valid in {\em any} instance that we might be in.
It is then clear that 
$$\K\Lambda^2(\sigma) ; \pi = \sigma ; \id{F{\U}} =\sigma$$
and hence that this definition fulfills the required properties.

Since we are interested in modeling IMLL2 (second order IMLL) 
we will refine our model with the notion of winning strategy, 
as explained in the previous section.

Firstly, we briefly indicate the additional structure required of 
a specification structure in order to get a model for IMLL2
in the refined category.

We assume that variable types are modeled by monotone functions 
$F:\G_{\U}\rightarrow\G_{\U}$ equipped with actions
$$F_{A}: PA\rightarrow P(FA)$$
for each $A\in\G_{\U}$.

Also there is an action:
$$\Pi_{F}:\one \rightarrow P(\Pi(F))$$
satisfying:

\begin{tabular}{cc}
$(\forall-\mbox{elim})$ &
$\Pi_F\{\pi_A\}\phi \ \ \ \ (A\in\G_{\U},\phi\in P(FA))$ \\
$(\forall-\mbox{intro})$ & 
$(\forall A\in\G_{\U}, \psi\in PA . \, \phi\{\sigma_A\}F_A(\psi))
\Rightarrow \phi\{\Lambda^2(\sigma)\}\Pi_F$.
\end{tabular}

Now in the case of the specification structure $W$ for winning strategies,
we define:
$$\Pi_F= \{s\in \P{\Pi(F)}^{\infty}\ |\ \forall A\in\G_{\U},
W\subseteq \P{A}^{\infty}. \, s\in \P{F(A)}^{\infty}\Rightarrow s\in 
F_A(W)\} .$$

\begin{exercise} \rm
Verify that this satisfies $(\forall$-intro$)$ and $(\forall$-elim$)$.
\end{exercise}

\vspace{1em}
\noindent
Thus we have a game semantics for IMLL2 in which terms denote
winning strategies. How good is this semantics? 
As a basic test, let us look at the type
\[ \forall X. \, X \multimap (X \multimap X) \]
which we write as
$$\forall X. X_1\multimap (X_2\multimap X_3)$$
using indices to refer to the occurrences of $X$.
What are the winning strategies for this type?
Note that the first move must be in $X_3$.
Because of the definition of $\Pi$, Player can only respond
by playing the same move in a negative occurrence of $X$, i.e $X_1$ or $X_2$.
Suppose Player responds in $X_2$: 
$$\begin{array}{cccc}
\forall X.\, X_1\multimap & (X_2 & \multimap & X_3)\\
& & & a \\
& a & &
\end{array}$$

At this point, by the switching condition Opponent must respond in
$X_2$, say with a move $b$ ; what can Player do next? If he were
playing as the term $\Lambda X.\lambda x,y:X.y$, then he should copy
$b$ back to $X_3$. However there is another possiblity (pointed out by
Sebastian Hunt): namely, Player can {\em play $a$ in $X_1$}, and
continue thereafter by playing copy-cat between $X_1$ and $X_3$. This
certainly yields a winning strategy, but does not correspond to the
denotation of any term.

To eliminate such undesirable possibilities, we introduce a constraint
on strategies. Recall from Exercise 1.10 that a strategy is {\em history-free} if its response at any point depends only on the last move by Opponent:
that is, if it satisfies:
\[
sab\in\sigma, ta\in P_A \THEN tab \in\sigma.
\]
The history-free strategies suffice to model the multiplicatives and
polymorphism, so we get a model $\Gwhf$ of IMLL2.

Now consider again the situation
$$\begin{array}{cccc}
\forall X. \, X_1\multimap & (X_2 & \multimap & X_3)\\
& & & a \\
& a & &\\
& b & &\\
\end{array}$$
Player can only respond to $b$ by copying $b$ into $X_3$ if he is
following a history-free strategy: the option of playing $a$ in $X_1$
is not open to him, because $a$ is not ``visible'' to him. Thus he can
only proceed by
$$\begin{array}{cccc}
\forall X. \, X_1\multimap & (X_2 & \multimap & X_3)\\
& & & a \\
& a & &\\
& b & &\\
& & & b
\end{array}$$
Moreover, Player must continue to play copy-cat between $X_2$ and
$X_3$ ever thereafter, since the information available to him at each
stage is only the move just played by Opponent.  

Note also that Player must play in the same way, regardless of which
move is initially made by Opponent. For example, suppose for a
contradiction that Player responded to $a_1$, by copying it to $X_1$,
and to $a_2$ by copying it to $X_2$. Now consider the situation:
$$\begin{array}{ccccccc}
\forall X. & X_1 & \multimap & (X_2 & \multimap & X_3)\\
&   &&   && a_1\\
& a_1  &&   && \\
& b_1  &&   && \\
&   &&   && b_1\\
&   &&   && a_2\\
&   && a_2  && \\
\end{array}$$
Since Player is following a history-free strategy, he must {\em
always} respond to $a_2$ by copying it to $X_2$; but the above position is
clearly not valid, since there is an instance $A$ with
$P_{A}={\mathsf Pref}\ASET{a_1b_1a_2}$ in which $a_2$ cannot be played as an
initial move.

Thus we conclude that for our test case the model $\Gwhf$ does indeed
have the required property that the only strategies for the game
\[
\forall X. \, X_1  \multimap  (X_2  \multimap  X_3)
\]
are the denotations of the terms:
\[
\begin{array}{ccc}
\Lambda X.\lambda x,y:X.x & & \Lambda X.\lambda x,y:X.y\\
\mbox{copycat between $X_1$ and $X_3$} & & 
\mbox{copycat between $X_2$ and $X_3$}.
\end{array}
\]

\begin{exercise}
Show that the only two strategies in $\Gwhf$ for the game
\[
\forall X. \, (X  \tensor X)  \multimap  (X  \tensor X)  
\]
are those corresponding to the identity and the twist map.
\end{exercise}

\begin{openproblem}
For which class of (closed) types of IMLL2 do we get a ``Full
Completeness'' result, i.e. that all strategies at that type in
$\Gwhf$ are definable in IMLL2?
\end{openproblem}

\section{Relational Parametricity}

In this section, we investigate how the notion of relational
parametricity can be adapted to the setting of games.

Firstly, we go back to the general level of Specification Structures.
We use some notions due to Andy Pitts (1996).

Given $\phi, \psi\in PA$, we define:
\[
\phi \leq \psi \;\; \DEFEQ \;\; \phi \ASET{\id{A}} \psi.
\]
This is always a preorder by \textbf{(ss1)} and \textbf{(ss2)}.  
Say that the specification structure $S$ is {\em posetal} 
if it is a partial order (i.e. antisymmetric). Now the notion
of meet of properties   $\AAND_{i\in I} \phi_i$ can be defined on $PA$.
Say that $S$ is {\em meet-closed} if it is posetal and each $PA$ has
all meets. 

Now we define a notion of {\em relations} on games. We shall focus on
binary relations. Say that $R$ is a relation from $A$ to $B$
(notation: $R\subseteq A\times B$) if $R$ is a non-empty subset
$R\subseteq P_A\times P_B$ satisfying:
\begin{itemize}
\item
$R(s,t) \THEN |s|=|t|$.
\item
$R(sa,tb) \THEN R(s,t)$.
\end{itemize}
(So $R$ is a length-preserving non-empty prefixed closed subset). 

We shall define a specification structure $R$ on the product category
$\G\times\G$ by taking $P(A,B)$ to be the set of relations
$R\subseteq A\times B$.  Given a relation $R\subseteq A\times B$, we lift
it to a relation $\MNTN{R}$ between strategies on $A$ and strategies on
$B$, by the following definition:
\[
\begin{array}{llllll}
\MNTN{R}(\sigma, \tau)  
& \Longleftrightarrow & \forall s\in\sigma, t\in\tau . \,  R(sa, ta') \\
&        & \begin{array}{ll}
           \THEN & [ (sa\in dom(\sigma) \IFF ta'\in dom(\tau))\\
                 & \ \wedge\  sab\in\sigma, ta'b'\in\tau \THEN R(sab, ta'b')]
\end{array}\\
\end{array}
\]
This definition is ``logical relations extended in time''; 
it relativizes the usual clause:
\[
R(x, y) \;\THEN\; [(fx{\downarrow} \  \Leftrightarrow \ gy{\downarrow}) 
\;\AND\;
(fx{\downarrow}, gy{\downarrow} \ \Rightarrow \ R(fx, gy))]
\]
to the context (previous history) $s$. It can also be seen as a form of
bisimulation:
\begin{quote}
``If $\sigma$ and $\tau$ reach related states at $P$'s turn to move,
then one has a response iff the other does, and the states after the
response are still related.''
\end{quote}
Also, if $R\subseteq A\times A'$ and $S\subseteq B\times B'$, then we
define:
\[
\begin{array}{llllll}
R\tensor_{\mbox{\tiny $(A,A'),(B,B')$}}S 
& = & \{ & (s,t)\in P_{A\tensor B}\times P_{A'\tensor B'}\;\;|\;\;\\
&   &    & R(s\RS A, t\RS A')\;\AND\; S(s\RS B, t\RS B')\\
&   &    &  \;\AND\; \mathtt{out}^\ast(s)= \mathtt{out}^\ast(t))\; \; \}
\end{array}
\]
where $\mathtt{out}: M_A+M_B\rightarrow \ASET{0,1}$ is given by:
\[
\mathtt{out}=[{\mathbf K}0, {\mathbf K}1]
\]
Similarly we define:
\[
\begin{array}{llllll}
R\multimap_{\mbox{\tiny $(A,A'),(B,B')$}}S 
& = & \{ & (s,t)\in P_{A\multimap B}\times P_{A'\multimap B'}\;\;|\;\\
&   &    & R(s\RS A, t\RS A'\;\AND\; S(s\RS B, t\RS B')\\
&   &    & \;\AND\; out^\ast(s)=out^\ast(t)) \; \; \}
\end{array}
\]

Now we define:
\[
R\ASET{(\sigma, \tau)}S
\;\DEFEQ\;
\MNTN{R\multimap S}(\sigma, \tau)
\]

\begin{proposition}
This is a specification structure in $\G\times\G$. In particular,
\[
R\ASET{(\sigma, \tau)} S,\; S\ASET{(\sigma', \tau')}T
\;\Longrightarrow\;
R\ASET{(\sigma;\sigma', \tau;\tau')}T
\]
\end{proposition}
\begin{centering}
$$
\xymatrix{
A\ar@{-}[dr]|-{R} \ar[rr]^{\sigma} & & B\ar@{-}[dr]|-{S} \ar[rr]^{\tau} 
& & C\ar@{-}[dr]|-{T}\\
& A'\ar[rr]^{\sigma'} & & B \ar[rr]^{\tau'} & & \; C'\\
&&b_1 \ar@{}[drrr]|-{\Longleftarrow} 
\ar@{-}[rr]\ar@{-}[dd] \ar@{-}[dr]|-{S} & &  c\ar@{-}[dr]|-{T}\\
&&& b'_1 \ar@{}[dl]|-{\Downarrow}\ar@{-}[dd]\ar@{-}[rr] & &c' \\
&&b_2 \ar@{-}[d] \ar@{-}[dr]|-{S} &\\
&&\ar@{.}[d] & b'_2\ar@{}[dl]|-{\Downarrow}\ar@{-}[d] \\
a\ar@{-}[rr]\ar@{-}[dr]|-{R}&&b_k\ar@{}[dl]|-{\Longleftarrow}\ar@{-}[dr]|-{S}& \ar@{.}[d]\\
&a'\ar@{-}[rr] &&b'_k
}
$$
\end{centering}

\begin{exercise}\rm
Prove this! (The above ``logical waterfall'' diagram gives the idea of the proof.)\\
\end{exercise}

We shall in fact be more interested in ``pulling back'' this
specification structure along the diagonal functor $\Delta:
\G\rightarrow \G\times\G$. That is, we are interested in the category
$\G_R$ with objects $(A, R)$ where $R\subseteq A\times A$ and morphisms
$\sigma: (A, R) \rightarrow (B, S)$ which are strategies $\sigma: A
\rightarrow B$ such that $\MNTN{R\multimap S}(\sigma, \sigma)$. We are also
interested in the category $\GwhfR$ where we combine the winning
strategy and relational structures, so that objects are $(A, W_A,
R_A)$, where $W_A$ is a set of designated winning plays, and
$R_A\subseteq A\times A$ is a relation and $\sigma: (A, W_A, R_A)$ is a
strategy $\sigma: A\rightarrow B$ such that 
$W_A \ASET{\sigma} W_B \AND R_A\ASET{\sigma}R_B$.

Now we build a model of IMLL2 by refining our previous model with this
specification structure $R$. A variable type will now be a monotone function
\[
F: (\G_{\U}, \ODR) \rightarrow (\G_{\U}, \ODR)
\]
with an action
\[
F_A: PA \rightarrow P(FA).
\]
We assume that the specification structure is monotone, in the sense that:
\[
A \ODR B \THEN PA \subseteq PB
\]
(this is easily seen to hold for $R$ and $W$), and that 
\begin{diagram}
PA                              & \rInto                & PB \\
\dTo^{F_A}                      &                       & \dTo_{F_B}\\
P(FA)                           & \rInto                & P(FB)\\
\end{diagram}
We also require that if $\phi\in PA, \psi\in PA, A \ODR A'$ and $B\ODR
B'$, then 
\[
\phi \ASET{F}_{A, B} \psi
\;\IFF\;
\phi \ASET{F}_{A', B'} \psi.
\]
We further assume that the specification structure is
meet-closed. Then we define:
\begin{equation}
\Pi_F \stackrel{df}{=} \bigwedge\ASET{F_A(\phi)\;|\; A\in\G_{\U}, \phi\in PA}\\
      = \bigwedge \ASET{F_A(\phi)\;|\; \phi\in P{\U}}
\end{equation}
(This latter equality holds because of the above monotonicity
properties). 

The fact that ($\forall$-intro) and ($\forall$-elim) are satisfied 
then automatically holds because of the definition of $\Pi_F$ as a meet.

To apply this construction to $R$, we must show that it is meet-closed.

Firstly, we characterise the partial order on properties in $R$.

\begin{proposition}
\[
\begin{array}{lllll}
R \le S \IFF & \;\; & R^{\mathrm{even}}(s, t)\AND S(sa, tb) & \THEN & R(sa, tb)\\
             & \AND & S^{\mathrm{odd}}(s, t)\AND R(sa, tb) & \THEN & S(sa, tb).
\end{array}
\]
\end{proposition}

We can read this as: at O-moves $S\subseteq R$ and at P-moves $R\subseteq
S$. 

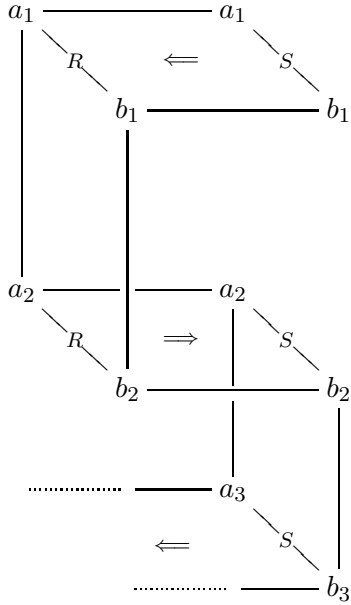
$$\xymatrix{
a_1 \ar@{}[drrr]|-{\Longleftarrow} \ar@{-}[rr]\ar@{-}[ddd] \ar@{-}[dr]|-{R} 
& &  a_1\ar@{-}[dr]|-{S}\\
& b_1 \ar@{-}[ddd]\ar@{-}[rr] & & b_1 \\
\\
a_2 \ar@{-}'[r][rr] \ar@{-}[dr]|-{R} \ar@{}[drrr]|-{\Longrightarrow} & &
a_2\ar@{-}'[d][dd] \ar@{-}[dr]|-{S}\\
& b_2\ar@{-}[rr] & & b_2\ar@{-}[dd]\\
\ar@{.}[r] &\ar@{-}[r] & a_3\ar@{-}[dr]|-{S}\\
&\ar@{}[ur]|-{\Longleftarrow} \ar@{.}[r] & \ar@{-}[r]& b_3}
$$

\begin{proposition}
$\bigwedge_{i\in I} R_i$  is defined inductively by:
\[
\begin{array}{lllllllll}
\AAND_{i\in I} R_i & = &      & \ASET{(\NUL, \NUL)}\\
                  &   & \cup & 
\ASET{(sa, ta')\; | & \; (s,t)\in \AND_{i\in I} R_i^{\mathrm{even}}\;\AND \;\\
                  &   &  &                    
                    & \exists i\in I. R_i(sa, ta')}\\
                  &   & \cup & 
\ASET{(sab, ta'b')\; | & \; (sa,ta')\in \AND_{i\in I} R_i^{\mathrm{odd}}
\; \AND \\
                  &   &  &                    
                    & \forall i\in I. R_i(sa, ta') \THEN R_i(sab, ta'b')}.
\end{array}
\]
\end{proposition}

Note the similarity between this definition and that of $P_{\Pi(F)}$,
which is in fact the unary case of the above, indexed over 
$P  {\NPRF} P_{F(U)}$.

\begin{exercise}\rm
\begin{enumerate}
\item
Verify these propositions.
\item
For the specification structure ${\cal W}$, show that:
\begin{itemize}
\item
$V\leq W \IFF V\subseteq W$.
\item
$\AAND_{i\in I} W_i = \bigcap_{i\in I} W_i$.
\end{itemize}
\end{enumerate}
\end{exercise}

Thus we obtain a model $\GwhfR$ of IMLL, incorporating both:
\begin{itemize}
\item
the refinement to winning strategies
\item
a notion of ``relational parametricity''.
\end{itemize}

\end{document}